\documentclass[aps,groupedaddress]{revtex4}
\usepackage{epsfig}
\usepackage{graphicx}
\usepackage[T1]{fontenc}
\usepackage{ae}
\usepackage{color}
\usepackage[latin1]{inputenc}
\usepackage{amssymb,amsbsy,amsmath}
\usepackage{bbm}
\begin{document}

%\begin{center}\today\end{center}

%%%%%%%%%%%%%%%%%%%%%%%%%%%%%%%%%%%%%%%%%%%%%%%%%%%%%%%%%%%%%%%%%%%%%%%%%%%%%

\title[Motion in a plate capacitor]
{On the motion of a point charge in a plate capacitor considering influence effects}

\author{Heinz-J\"urgen Schmidt and Thomas Br\"ocker
}
\affiliation{ Universit\"at Osnabr\"uck,
Fachbereich Mathematik, Informatik und Physik,
 D - 49069 Osnabr\"uck, Germany
}

%\tableofcontents

\begin{abstract}
A point charge between the plates of a capacitor generates an influence charge distribution
on the plates that modify the electric field acting upon the point charge. This effect
is described by the well-known Dirichlet Green's function for the two parallel conducting plate problem
for which we derive an infinite mirror charge series representation. At the line
perpendicular to the plates and passing through
the point charge this Green's function and hence the total force can be explicitly evaluated
in terms of the psi function.
For the motion of the point charge we develop an analytical approximation and compare it with the
numerical integration of the exact equations of motion. The correction due to influence effects
is shown to be of order $O(-\lambda \log \lambda)$ where $\lambda$ denotes the relative strength
of the Green's function compared with the pure capacitor potential.
\end{abstract}

\maketitle

%%%%%%%%%%%%%%%%%%%%%%%%%%%%%%%%%%%%%%%%%%%%%%%%%%%%%%%%%%%%%%%%%%%%%%%%%%%%%%%%%%%
\section{Introduction}\label{sec:I}
%%%%%%%%%%%%%%%%%%%%%%%%%%%%%%%%%%%%%%%%%%%%%%%%%%%%%%%%%%%%%%%%%%%%%%%%%%%%%%%%%%%

Many supposedly simple problems in physics turn out to be not so simple on closer inspection.
As an example of this we look at the motion of a point charge
in the field of a plate capacitor.
At first glance, this is an example of motion in a constant force field
that could be covered in physics lessons at school.
In this simplified view, however, the sources of the field,
i.~e.~the surface charges on the capacitor plates, are assumed to be fixed.
In reality, however, the point charge acts back on the sources and generates ``influence charges"
on both plates, which have an opposite sign relative to the point charge.
These influence charges change the field, and hence also the acceleration
of the point charge, and in turn depend on the location of the point charge.
The overall result is a complex interplay between the motion of the point charge
and influence effects, in which the distinction between cause and effect becomes blurred.

An intermediate step in the theoretical treatment of the problem outlined
is the calculation of the (Dirichlet) Green's function $G({\mathbf r},{\mathbf r}_0)$
for the plate capacitor. This is essentially the electric potential of a point charge
at the position ${\mathbf r}_0$ between the plates with the Dirichlet boundary condition
that $G({\mathbf r},{\mathbf r}_0)$ vanishes if ${\mathbf r}$ lies on the surface of the capacitor plates.
From this, the potential energy for the motion of the point charge can be obtained
by subtracting its Coulomb potential from $G({\mathbf r},{\mathbf r}_0)$,
adding the pure capacitor potential and setting ${\mathbf r}={\mathbf r}_0$ in the result.
Of course, we assume that the influenced surface charges react instantaneously
to the motion of the point charge without any delay.
The Green's function for two parallel conducting plates is known; it is given in the
literature as a double series expansion according to cylindrically symmetric
eigenfunctions of the Laplace operator, see \cite{J21}, Problem 3.17a,
or as a mixed series/integral representation, see \cite{J21}, Problem 3.17b.
An early source is \cite{D1900}. \S 19.

However, the double series representation is less suitable for our problem, see Appendix \ref{sec:CO}, 
and also the series/integral representation appears opaque at first sight.
We therefore propose a different way of calculating $G({\mathbf r},{\mathbf r}_0)$,
which is based on the idea of mirror charges.
The approach of calculating the potential of a point charge in a plate capacitor by means of
infinitely any mirror charges is occasionally
discussed on the internet and in lecture notes, but to our knowledge has rarely been carried out explicitly.
Exceptions are \cite{M20}, where the special case is considered that the point charge is located exactly in the middle
between the plates, and \cite{N82}.
It is well known that, in the presence
of only one conducting plate, the Green's function can be constructed
by assuming, in addition to the point charge
$q$ at a distance $x$ from the plate, a fictitious ``mirror charge"
$-q$ at the same distance $x$ behind the plate and defining $G({\mathbf r},{\mathbf r}_0)$
as the sum of the potentials of both charges. If two plates are present,
one can work with two ``primary" mirror charges accordingly.
Each of these mirror charges ensures a constant potential on the plate assigned to it,
but distorts the potential on the other plate.
To compensate for this, additional ``secondary" mirror charges can be introduced,
but then the analogous problem will arise, albeit to a lesser extent.
Overall, the Green's function results as an (initially formal) series of
Coulomb potentials of an infinite number of mirror charges at ever greater distances
from the plates, in addition to the Coulomb potential of the actual point charge,
see section \ref{sec:IS}.
The convergence of this series representation is proven in the appendix \ref{sec:CS}.
Similar approaches involving infinite many mirror charges have been applied to the
problem of a point charge on the axis of a conducting cylinder \cite{M20} or
to find the self- and mutual capacitances for two spheres \cite{S89}, section 5.08 and 5.081.

We have thus obtained an alternative series representation of Green's function
for the plate capacitor, which is somewhat simpler than the representations
known from the literature and whose terms have a clear meaning as potentials
of mirror charges. The motion of the point charge can be restricted
to a straight line $L$ perpendicular to the plates, because the motion
perpendicular to it is rectilinear and uniform. We will refer to this straight line $L$,
on which the mirror charges also lie, as the ``central line" in this paper
and choose it as the $\xi$ axis in a rectangular coordinate system $(\xi, \eta, \zeta)$.
As explained above, the special values $G(\xi,\xi_0)$ of the restricted Green's function are
required for the motion of the point charge on the central line.
It now turns out that the series representation for $G(\xi,\xi_0)$
can be explicitly evaluated in terms of the so-called
psi function $\psi(z)= \Gamma'(z)/ \Gamma(z)$, see section \ref{sec:EG}.
An alternative derivation using the mixed series/integral representation of
the Green's function is given in Appendix \ref{sec:CO}.
This also allows the force on the point charge to be calculated and the
corresponding equation of motion for $x(t)$ to be numerically integrated.
In principle, we have thus solved the problem that we set ourselves in this paper.

In the following, we investigate the possibilities of analytical approximations
for $x(t)$, which are offered by our approach. In the vicinity of the left-hand plate,
the influence of all mirror charges can be approximated by that of the
single primary mirror charge, which is located at a distance $x(t)$
behind the plate, see subsection \ref{sec:MA}. This leads to the motion in a potential
of the form $V(x)=-q U \frac{x}{d}-\frac{q^2}{4\pi \epsilon_0}\frac{1}{2x}$,
or, in dimensionless quantities, $V(x) =-x-\frac{\lambda}{x}$.
This simple potential, superposition of a linear and a
Coulomb potential, is known in the literature as ``Cornell potential"
and is used in connection with confinement in quantum chromodynamics, see \cite{S83}.
Another application of this potential is the Stark effect, see \cite{Cetal95}.
It turns out that the one-dimensional motion in the Cornell potential
can be described by Legendre elliptic integrals of the 1st and 2nd kind for the inverse function $t(x)$,
see subsection \ref{sec:MCL} and Appendix \ref{sec:AE}. This also applies to the motion near the right-hand plate
in the approximated potential $V(x))=-x-\frac{\lambda}{1-x}$, see subsection \ref{sec:MCR}.
We therefore divide the range $0\le x \le 1$ between the plates into three equal parts
and approximate the exact potential in the middle range $1/3 \le x \le 2/3$
by its second-order Taylor expansion at the point $x=1/2$.
The approximated equation of motion can also be solved exactly in this range,
see subsection \ref{sec:MCM}.
The quality of the analytical approximation just described for the motion $x(t)$
of the point charge is excellent for the parameters investigated here,
see subsection \ref{sec:MCNA}.

The relative strength of the influence effects is described by the dimensionless parameter $\lambda$.
For $\lambda=0$, the (dimensionless) transit time $T(\lambda)$ of the point charge between the plates
assumes the simple value $T(0)=\sqrt{2} \left( \sqrt{E+1}-\sqrt{E} \right)$,
where $E$ denotes the total energy. At first glance one would expect
that the influence correction $\delta T= T(0)-T(\lambda)$ approaches zero with the order $O(\lambda)$.
In fact, however, the portions of the potential of the form $-\frac{\lambda}{x}$ and $-\frac{\lambda}{1-x}$,
which describe the influence of the primary mirror charges, are arbitrarily large for $x\to 0$, resp.~$x\to 1$,
as long as $\lambda>0$. It is therefore plausible that $\delta T$ approaches zero
more slowly than with the order $O(\lambda)$. Numerical and analytical investigations
rather show a $\lambda$-dependence of $\delta T$ of the order $O(-\lambda\,\log\,\lambda)$,
see section \ref{sec:IC}.

After the successful theoretical treatment of the motion of a point charge in a plate capacitor,
the question of possible experimental applications arises.
In the last decades, experiments with a so-called ``electrostatic pendulum" have been reported,
as well as approaches to their theoretical description,
see \cite{ML72,K75,P02,S12,FMG14,Getal20}.
This is a charged hollow sphere that oscillates between the plates of a capacitor
and is recharged after contact with one plate at a time.
The complete description of this experiment is difficult
because it would have to include influence effects on the capacitor plates
as well as on the sphere. The present work can be seen as a theoretical description
of a limiting case of these pendulum experiments in which,
firstly, the pendulum length is large compared to the distance of the plates
(and therefore the motion is almost rectilinear) and, secondly,
the radius of the charged sphere is so small that it can be described as a point charge.
We will return to this aspect of our work in the Summary and Outlook section \ref{sec:SO}.

%%%%%%%%%%%%%%%%%%%%%%%%%%%%%%%%%%%%%%%%%%%%%%%%%%%%%%%%%%%%%%%%%%%%%%%%%%%%%%%%
\section{Infinite sequence of mirror charges}\label{sec:IS}
%%%%%%%%%%%%%%%%%%%%%%%%%%%%%%%%%%%%%%%%%%%%%%%%%%%%%%%%%%%%%%%%%%%%%%%%%%%%%%%%%%%

\begin{figure}[ht]
  \centering
    \includegraphics[width=1.0\linewidth]{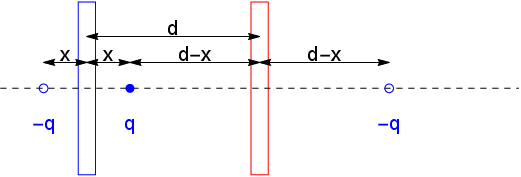}
  \caption{
  Sketch of the problem considered in this paper: We show two capacitor plates at a distance $d$.
  A particle with charge $q$ moving between the plates has at momentary distance $x$ from the left plate,
  and at distance $d-x$ from the right plate. We consider a ``primary" left mirror charge $-q$
  at distance $x$ from the left plate and another primary right mirror charge $-q$
  at a distance $d-x$ from the right-hand plate.
  Both primary mirror charges generate series of further ``secondary" mirror charges
  with alternating signs, which are located at the central line (dashed black line)
  to the right or left of the capacitor plates.
  }
  \label{FIGPK}
\end{figure}

\begin{figure}[ht]
  \centering
    \includegraphics[width=1.0\linewidth]{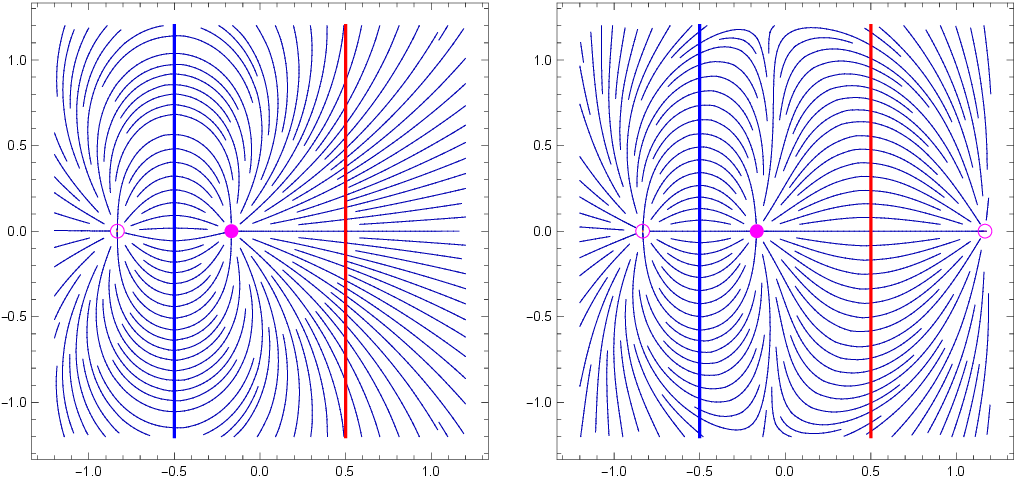}
  \caption{
  Left panel: Sketch of the field generated by a point charge (magenta dot) and a left
  mirror charge (magenta circle). The field lines are perpendicular on the left plate (blue line).\\
  Right panel: Sketch of the field  generated by a point charge and two ``primary" mirror charges (magenta circles).
  The field lines are approximately perpendicular on the plates (blue and red line) only in a central region close to the
  central line joining the charges.
  }
  \label{FIGMC12}
\end{figure}

The potential of a point charge $q$ located at a distance $x$ in front of an infinitely
extended conductive plate can be calculated using the well-known mirror charge method.
This consists of assuming a fictitious mirror charge with opposite sign $-q$
at the same distance $x$ behind the plate and adding the fields
of the point charge and the mirror charge, see Figure \ref{FIGPK}.
The field lines of the resulting field are then perpendicular
to the surface of the plate, which therefore forms an equipotential surface of the resulting potential,
see Figure \ref{FIGMC12}, left panel.
If a second conductive parallel plate is added so that the point charge is located
between the plates of a capacitor with the distance $d$,
it is obvious to also introduce a second mirror charge $-q$ at a distance
$d-x$ behind the second plate, see Figure \ref{FIGPK}.
These two mirror charges are referred to as ``primary" mirror charges.

\begin{figure}[ht]
  \centering
    \includegraphics[width=1.0\linewidth]{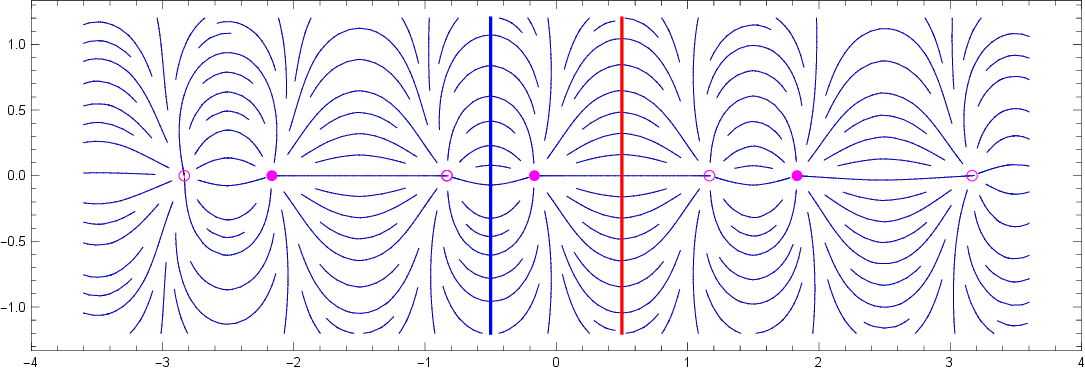}
  \caption{
  Sketch of the field generated by a point charge, two ``primary" mirror charges
  and $8$ further ``secondary" mirror charges of alternate signs,
  of which only $4$ are visible. The total field is vertical on both plates to a very good approximation.
  The actual field of the point charge will be obtained by the limit of the infinitely many
  mirror charges, restricted to the space between the plates.
  The fictitious field outside this space is shown here for illustrative purposes only.
  }
  \label{FIGMC3}
\end{figure}

However, the field lines of the system, which consists of a point charge
and two mirror charges, are not exactly perpendicular to the surfaces of the
two plates, see Figure \ref{FIGMC12}, right panel,
because the field of the left mirror charge is not perpendicular
to the right plate and, conversely, the field of the right mirror charge
is not perpendicular to the left plate. To further approximate a solution
to the electrostatic problem, additional ``secondary" mirror charges
of the primary mirror charges can be introduced.
But then the analogous problem arises again, albeit to a lesser extent.
It is therefore expected that the exact potential of a point charge between two
parallel conducting plates can only be obtained by adding the potentials of
an infinite number of mirror charges at ever greater distances from the plates,
see figure \ref{FIGMC3}.

\begin{figure}[ht]
  \centering
    \includegraphics[width=0.7\linewidth]{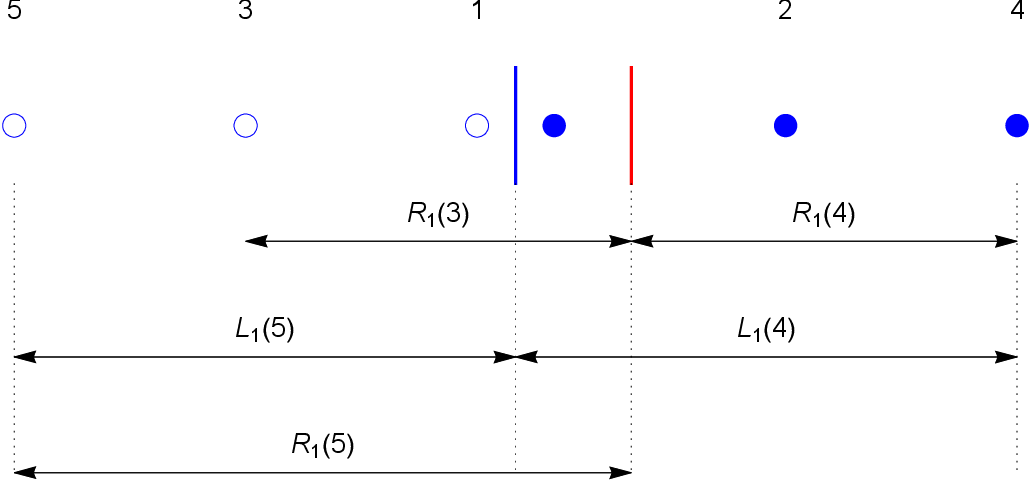}
  \caption{
  Sketch of the first five mirror charges $q_n, n=1,\ldots,5$ of the first generation.
  The first row contains their numbers $n=1,\ldots,5$.
  The odd charges $q_1,q_3,q_5,\ldots$ are reflected at the right plate of the capacitor
  which leads to the even charges $q_2,q_4,\ldots$.
  The corresponding distances to the right plate are therefore the same, for example $R_1(4)=R_1(3)$.
  The distance of the even charges to the left-hand plate is increased by $d$, for example $L_1(4)=R_1(4)+d=R_1(3)+d$.
  The even charges $q_2,q_4,\ldots$ are reflected at the left plate, resulting in the odd charges $q_3,q_5,\ldots$.
  Therefore, the corresponding distances to the left plate are equal, for example $L_1(5)=L_1(4)$.
  In addition, the distance of the odd charges to the right plate is increased by $d$, for example $R_1(5)=L_1(5)+d=L_1(4)+d$.
  This illustrates the recurrence relations (\ref{L1n}) and (\ref{R1n}) for the first generation of mirror charges.
  }
  \label{FIGG1}
\end{figure}

\begin{figure}[ht]
  \centering
    \includegraphics[width=0.7\linewidth]{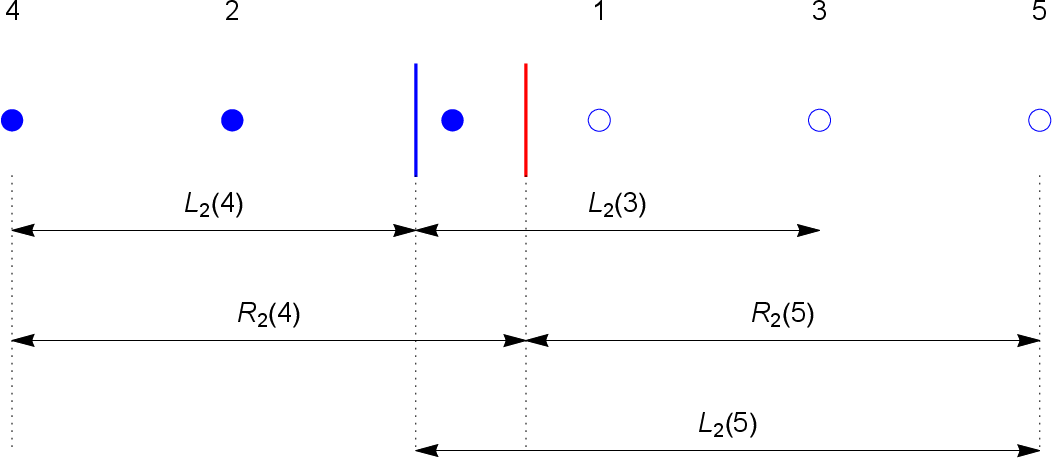}
  \caption{
  Sketch of the first five mirror charges $q_n, n=1,\ldots,5$ of the second generation.
  The first row contains their numbers $n=1,\ldots,5$.
  The odd charges $q_1,q_3,q_5,\ldots$ are reflected at the left plate of the capacitor
  which leads to the even charges $q_2,q_4,\ldots$.
  The corresponding distances to the left plate are therefore the same, for example $L_2(4)=L_2(3)$.
  The distance of the even charges to the right plate is increased by $d$, for example $R_2(4)=L_2(4)+d=L_2(3)+d$.
  The even charges $q_2,q_4,\ldots$ are reflected at the right plate, resulting in the odd charges $q_3,q_5,\ldots$.
  Therefore, the corresponding distances to the right plate are equal, for example $R_2(5)=R_2(4)$.
  In addition, the distance of the odd charges to the left plate is increased by $d$, for example $L_2(5)=R_2(5)+d=R_2(4)+d$.
  This illustrates the recurrence relations (\ref{L2n}) and (\ref{R2n}) for the second generation of mirror charges.
  }
  \label{FIGG2}
\end{figure}

To calculate the positions of the various mirror charges it is advisable to divide
the set of mirror charges into two ``generations" depending on whether we start with the left
or the right primary mirror charge. Let $R_1(n)$  denote the distance of the $n$th mirror
charge $q_n, n=1,2,\ldots$ of the first generation to the right plate, analogously $L_1(n)$ for the
distance to the left plate, and $R_2(n)$  and $L_2(n)$ for the mirror charges of the second generation, see
Figures \ref{FIGG1} and  \ref{FIGG2}. Then the distances of the primary mirror charges,
i.~e., for $n=1$,  satisfy
\begin{eqnarray}\label{RL1}
 R_1(1)&=&d+x,\quad L_1(1)=x,\\
 \label{RL2}
 L_2(1)&=&2 d-x,\quad R_2(1)=d-x
 \;,
 \end{eqnarray}
see Figure \ref{FIGPK}.
The mirror charges have alternating signs $(-1)^n q$ for both generations
so that, assuming $q>0$, all negative charges are at the left hand side of the left plate
for the first generation, and at the right hand side of the right plate
for the second generation, see the Figures \ref{FIGG1} and  \ref{FIGG2}.
Moreover, the four sequences $n\mapsto L_i(n), R_i(n), i=1,2,$ satisfy
the recursion relations
\begin{eqnarray}
\label{L1n}
 L_1(n+1) &=& \left\{
\begin{array}{r@{\quad : \quad}l}
   L_1(n) & n \mbox{ even} \\
   R_1(n)+d & n \mbox{ odd}
 \end{array}
 \right. \\
 \label{R1n}
 R_1(n+1) &=& \left\{
\begin{array}{r@{\quad : \quad}l}
   L_1(n)+d & n \mbox{ even} \\
   R_1(n) & n \mbox{ odd}
 \end{array}
 \right. \\
 \label{L2n}
 L_2(n+1) &=& \left\{
\begin{array}{r@{\quad : \quad}l}
   R_2(n)+d & n \mbox{ even} \\
   L_2(n) & n \mbox{ odd}
 \end{array}
 \right. \\
 \label{R2n}
 R_2(n+1) &=& \left\{
\begin{array}{r@{\quad : \quad}l}
   R_2(n) & n \mbox{ even} \\
  L_2(n)+d & n \mbox{ odd}
 \end{array}
 \right.
 \;.
\end{eqnarray}
These relations can be obtained by generalizing the relations considered in the Figures
\ref{FIGG1} and \ref{FIGG2}.
The initial conditions (\ref{RL1}),(\ref{RL2})
and the recursion relations (\ref{L1n}) -(\ref{R2n}) lead to the following
sequences with typical pairwise constant sequence members:
\begin{eqnarray}
\label{SL1}
 L_1 &=&(x, 2 d + x, 2 d + x, 4 d + x, 4 d + x,\ldots)\;,\\
 \label{SR1}
 R_1 &=&(d + x, d + x, 3 d + x, 3 d + x, 5 d + x, 5 d + x,\ldots)\;,\\
 \label{SL2}
 L_2 &=&(2d - x, 2d - x, 4 d - x, 4 d - x, 6 d - x, 6 d - x,\ldots)\;,\\
 \label{SR2}
 R_2 &=&(d - x, 3 d - x, 3 d - x, 5 d - x, 5 d - x,\ldots)
 \;.
\end{eqnarray}
Their explicit representation can be given in the following form:
\begin{eqnarray}
\label{DefL1}
  L_1(n) &=& 2 \left\lfloor\frac{n}{2}\right\rfloor d+x\;, \\
  \label{DefR1}
  R_1(n) &=& \left( 2 \left\lfloor\frac{n+1}{2}\right\rfloor -1\right) d+x\;, \\
  \label{DefL2}
  L_2(n) &=& 2 \left\lfloor\frac{n+1}{2}\right\rfloor d-x\;, \\
  \label{DefR2}
  R_2(n) &=& \left( 2 \left\lfloor\frac{n}{2}\right\rfloor +1\right) d-x
  \;,
\end{eqnarray}
for $n=1,2,\ldots$. The {\bf proof} is straightforward. In order to give an example
we will only prove that (\ref{L1n}) is satisfied by (\ref{DefL1}) and (\ref{DefR1}):
First, let $n$ be even. Then
\begin{equation}\label{neven}
 L_1(n+1)\stackrel{(\ref{DefL1})}{=}2 \left\lfloor\frac{n+1}{2}\right\rfloor d+x
 \stackrel{(n \mbox{\scriptsize{ even}})}{=}2 \left\lfloor\frac{n}{2}\right\rfloor d+x \stackrel{(\ref{DefL1})}{=}L_1(n)
 \;.
\end{equation}
Next, let $n$ be odd. Then
\begin{equation}\label{nodd}
  L_1(n+1)\stackrel{(\ref{DefL1})}{=}2 \left\lfloor\frac{n+1}{2}\right\rfloor d+x
  =  \left(2 \left\lfloor\frac{n+1}{2}\right\rfloor -1 \right)d+x+d
  \stackrel{(\ref{DefR1})}{=}R_1(n)+d
  \;.
\end{equation}
This completes the proof of (\ref{L1n}).   \hfill$\Box$\\

To determine the coordinates of the mirror charges
we define the central line connecting the mirror charges as a
$\xi$-axis with the origin in the center exactly between the two plates,
so that the original point charge has the coordinate
\begin{equation}\label{xi0}
 \xi_0 = x- \frac{d}{2}
 \;.
\end{equation}
Then a short calculation gives the $\xi$-coordinates of the mirror
charges of the two generations in the following form:
\begin{eqnarray}
\label{X1}
  X_1(n) &=& (-1)^n \left( \left( n-\frac{1}{2}\right) d +x\right)\;, \\
\label{X2}
  X_2(n) &=& (-1)^n \left(- \left( n+\frac{1}{2}\right) d +x\right)
   \;,
\end{eqnarray}
for $n=1,2,\ldots$.

We extend the $\xi$-axis to a $3$-dimensional Cartesian coordinate system with coordinates
${\mathbf r}=(\xi,\eta,\zeta)$ such that the mirror charges obtain the coordinates
\begin{eqnarray}
\label{CR1}
 {\mathbf R}_1(n) &=& \left(X_1(n),0,0 \right) \\
 \label{CR2}
 {\mathbf R}_2(n) &=& \left(X_2(n),0,0 \right)
\end{eqnarray}
for $n=1,2,\ldots$ and the point charge has the coordinates
\begin{equation}\label{r0}
{\mathbf r}_0=(\xi_0,0,0)
\;.
\end{equation}

The electrical potential of the point charge in the plate capacitor is formally given by the following infinite sum
of Coulomb potentials
\begin{equation}\label{Phi}
 \Phi({\mathbf r},{\mathbf r}_0)= \frac{q}{4\pi \epsilon_0}\left( \sum_{i=1,2}\sum_{n=1}^\infty
 \frac{(-1)^n}{\left|{\mathbf r}-{\mathbf R}_i(n) \right|} + \frac{1}{\left|{\mathbf r}-{\mathbf r}_0 \right|}
 \right)
\end{equation}
We have labeled this solution as ``formal'' since questions of convergence of the
series have not yet been considered. For this see Appendix \ref{sec:CS}.

Next we address the question whether the series representation (\ref{Phi})
can be considered as the Dirichlet Green's function of the plate capacitor problem.
For this it is required as a boundary condition that the potential assumes the value zero at both plates.
This would then automatically entail that the field lines of the corresponding electrical field
are perpendicular to the surfaces of the plates.
Anticipating the considerations in Appendix \ref{sec:CS} we arrange the contributions to (\ref{Phi})
into pairs. For the second generation, it is obvious that the pairs of mirror charges numbered
$(1,2),\, (3,4),\,\ldots$ are symmetric with respect to the left plate, see Figure \ref{FIGG2},
and, since they have opposite charges, that the sum of the potentials of each pair vanishes at the left plate.
For the first generation the situation is analogous for the pairs with numbers $(2,3),\,(4,5),\ldots$,
see Figure \ref{FIGG1}, such that the contribution of these pairs also vanishes at the left plate.
The remaining terms correspond to the primary mirror charge with number $n=1$ and to the point charge.
But also their potentials add to zero at the left plate.
The argumentation for the vanishing of (\ref{Phi}) at the right-hand plate is completely analogous.
This shows that, upon arranging the series (\ref{Phi}) pairwise in the manner indicated above,
it represents a potential that satisfies the requirement of a Dirichlet Green's function,
i.~e., $G({\mathbf r},{\mathbf r}_0)=\Phi({\mathbf r},{\mathbf r}_0)$ if setting $\frac{q}{4\pi \epsilon_0}=1$.

%%%%%%%%%%%%%%%%%%%%%%%%%%%%%%%%%%%%%%%%%%%%%%%%%%%%%%%%%%%%%%%%%%%%%%%%%%%%%%%%%%%
\section{Evaluation of the Green's function at the central line}\label{sec:EG}
%%%%%%%%%%%%%%%%%%%%%%%%%%%%%%%%%%%%%%%%%%%%%%%%%%%%%%%%%%%%%%%%%%%%%%%%%%%%%%%%%%%

We will explicitly evaluate the series representation (\ref{Phi}) of $\Phi({\mathbf r},{\mathbf r}_0)$
for the special values of ${\mathbf r}=(\xi,0,0)$, that is, at the central line joining the mirror charges.
Based on the considerations in the Appendix \ref{sec:CS} on the conditional convergence
of the series  (\ref{Phi}), we will arrange the terms for both generations of mirror charges pairwise.

Starting with the first generation and recalling the definitions (\ref{xi0}), (\ref{CR1}) and (\ref{r0})
we obtain for even $n=2m$
\begin{equation}\label{dneven}
 \left| {\mathbf R}_1(2m) -{\mathbf r}\right| = 2 m d +\xi_0 -\xi
 \;,
\end{equation}
while for odd $n=2m-1$
\begin{equation}\label{dnodd}
 \left| {\mathbf R}_1(2m-1) -{\mathbf r}\right| = (2 m-1) d +\xi_0 +\xi
 \;.
\end{equation}
This entails
\begin{equation}\label{together1}
\frac{1}{ \left| {\mathbf R}_1(2m) -{\mathbf r}\right|} - \frac{1}{ \left| {\mathbf R}_1(2m-1) -{\mathbf r}\right|}
=\frac{2 \xi -d}{(2 m d-\xi +\xi_0) ( (2 m-1)d+\xi +\xi_0)}
\;.
\end{equation}
The infinite sum of these terms can be explicitly calculated and gives the result
\begin{equation}\label{sumpairsgen1}
 \varphi_1:= \sum_{m=1}^{\infty}
 \frac{1}{ \left| {\mathbf R}_1(2m) -{\mathbf r}\right|} - \frac{1}{ \left| {\mathbf R}_1(2m-1) -{\mathbf r}\right|}
 =
 \frac{1}{2d}\,\left(-\psi\left(\frac{2 d-\xi +\xi _0}{2 d}\right)+
 \psi\left(\frac{d+\xi +\xi _0}{2 d}\right)\right)
 \;,
\end{equation}
where $\psi(z)= {\Gamma'(z)}/{\Gamma(z)}$ denotes the psi function, see \cite{NIST}, eq.~5.2.2.~,
which is a special case of the polygamma functions $\psi^{(n)}(z)$, see \cite{NIST},\,\S\, 5.15.

The result (\ref{sumpairsgen1}) has been derived with the aid of computer algebra software;
for similar series evaluations see \cite{NIST}, chapter 5.7.

Analogously, we obtain for the second generation
\begin{equation}\label{together2}
\frac{1}{ \left| {\mathbf R}_2(2m) -{\mathbf r}\right|} - \frac{1}{ \left| {\mathbf R}_2(2m-1) -{\mathbf r}\right|}=
\frac{1}{2 d m+\xi -\xi _0}-\frac{1}{d (2 m-1)-\xi -\xi _0}=
\frac{d+2 \xi }{\left(2 d m+\xi -\xi _0\right) \left(-2 d m+d+\xi +\xi _0\right)}
\;,
\end{equation}
and the infinite sum of of these terms is given by
\begin{equation}\label{sumpairsgen2}
 \varphi_2:= \sum_{m=1}^{\infty}
 \frac{1}{ \left| {\mathbf R}_2(2m) -{\mathbf r}\right|} - \frac{1}{ \left| {\mathbf R}_2(2m-1) -{\mathbf r}\right|}
 =
 \frac{1}{2d}\,\left(\psi\left(\frac{d-\xi-\xi_0}{2 d}\right)-\psi\left(\frac{2 d+\xi-\xi_0}{2d}\right)\right)
 \;.
\end{equation}

\begin{figure}[ht]
  \centering
    \includegraphics[width=0.7\linewidth]{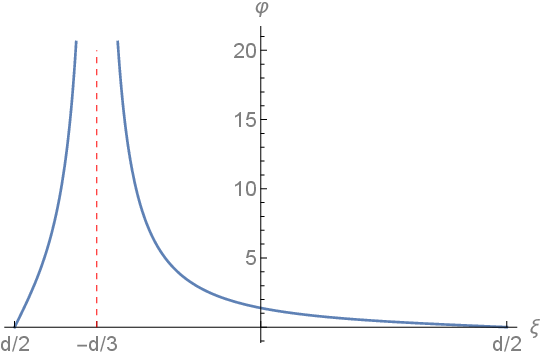}
  \caption{
 The potential of a point charge and an infinite number of mirror charges $\varphi(\xi,\xi_0)$,
 restricted to the $\xi$-axis according to (\ref{phi}) (arbitrary units).
 The coordinate of the point charge is chosen as $\xi_0=-d/3$ where $\varphi(\xi,\xi_0)$ diverges according
 to the contribution of the Coulomb potential. Remarkably, $\varphi(\xi,\xi_0)$ vanishes for $\xi \to \pm d/2$
 as it is required for a Dirichlet Green's function.
  }
  \label{FIGPHI}
\end{figure}

Finally, we add the contribution of both generations and the Coulomb potential of the point charge to obtain
\begin{equation}\label{phi}
 \varphi(\xi,\xi_0) := \varphi_1 +\varphi_2 +\frac{1}{\left| \xi-\xi_0\right|}
 \;,
 \end{equation}
 which, up to the pre-factor $\frac{q}{4\pi \epsilon_0}$, represents an explicit evaluation
 of the series representation (\ref{Phi}) restricted to the $\xi$-axis.

 The limit value of the potential (\ref{Phi}) for $\xi\to \pm d/2$ has already be shown to vanish
 in Section \ref{sec:IS}. As a test one can check that this also applies for its
 evaluation (\ref{phi}) at the $\xi$-axis. Interestingly, this test has to use the
 recurrence equation $\psi(z+1) =\psi(z) + 1/z$ for the psi function, see \cite{NIST}, eq.~5.5.2.

%%%%%%%%%%%%%%%%%%%%%%%%%%%%%%%%%%%%%%%%%%%%%%%%%%%%%%%%%%%%%%%%%%%%%%%%%%%%%%%%%%%
\section{Motion of the point charge in the plate capacitor}\label{sec:M}
%%%%%%%%%%%%%%%%%%%%%%%%%%%%%%%%%%%%%%%%%%%%%%%%%%%%%%%%%%%%%%%%%%%%%%%%%%%%%%%%%%%
\subsection{Approximation of the potential energy}\label{sec:MA}
%%%%%%%%%%%%%%%%%%%%%%%%%%%%%%%%%%%%%%%%%%%%%%%%%%%%%%%%%%%%%%%%%%%%%%%%%%%%%%%%%%%

At this point, it is advisable to introduce dimensionless variables.
We therefore consider the following quantities with their problem-oriented units:\\

\begin{tabular}{|c|c|c|c|c|}
\hline
Length & Mass & Charge & Energy & Time \\
\hline
 $d$ & ${\sf m}>0$ & $q>0$ & $q\,U$ &$\sqrt{\frac{{\sf m}}{q U}}\,d$ \\
 \hline
\end{tabular}
\\*[3mm]
Here ${\sf m}$ denotes the mass of the point charge and $U>0$ the voltage between the
plates of the capacitor, assuming that the left plate is positively charged
and the right one negatively. We will use the same letters to designate the
dimensionless variables as before, without the risk of confusion.
The potential energy of a pair of mirror charges with distance $2d$ divided by
the energy unit $q\,U$ will define a dimensionless parameter $\lambda$ typical for the
relative strength of influence effects:
\begin{equation}\label{deflambda}
 \lambda:= \frac{q^2}{4\pi \epsilon_0 2 d}\frac{1}{q U}=\frac{q}{8\pi \epsilon_0 U d}
 \;.
\end{equation}

For the potential energy of the point charge corresponding to the mirror charges we obtain
\begin{equation}\label{pot1}
V_1 = \left.2\lambda \left( \varphi_1 +\varphi_2\right)\right|_{\xi=\xi_0}
\stackrel{(\ref{sumpairsgen1},\ref{sumpairsgen2})}{=}
\lambda\left( \psi\left(\frac{1}{2}-\xi_0\right)+\psi\left(\xi_0+\frac{1}{2}\right)+2 \gamma \right)
\stackrel{(\ref{xi0})}{=}
\lambda(\psi(1-x)+\psi(x)+2 \gamma )
\;,
\end{equation}
where $\gamma= 0.57721\ldots$ denotes the Euler-Mascheroni constant.

For the motion of the point charge the potential energy of the plate capacitor
without influence effects has to be added which gives the total potential energy
\begin{equation}\label{totalpot}
V(x) = V_1 - x = \lambda(\psi(1-x)+\psi(x)+2 \gamma ) -x
\;.
\end{equation}
From this the total force $F(x)$ upon the point charge can be obtained by
\begin{equation}\label{totalF}
 F(x)=-\frac{d}{dx}V(x)=1+\lambda\left(  \psi ^{(1)}(1-x)- \psi ^{(1)}(x)\right)
 \;.
\end{equation}

\begin{figure}[ht]
  \centering
    \includegraphics[width=0.7\linewidth]{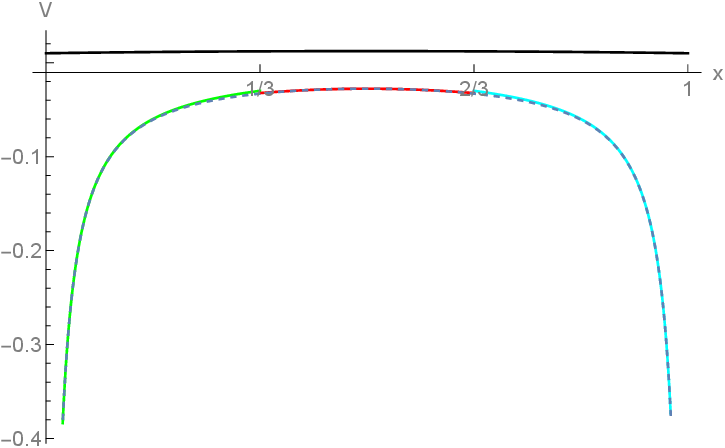}
  \caption{
 The potential energy $V_1(x)$ of a point charge due to an infinite number of mirror charges (dashed curve)
 and its analytical approximation $V_{app}(x)$ consisting of three parts according to (\ref{analapp})
 (green, red and cyan curves). Moreover, we show the potential energy $V_2(x)$ only due to the
 secondary mirror charges according to (\ref{defV2}), see the black curve.
 The parameter $\lambda$ has been chosen as $\lambda=0.01$.
  }
  \label{FIGAPP}
\end{figure}

It is obvious that the corresponding equation of motion cannot be solved in closed form
but only numerically. Therefore it is sensible to devise an analytical approximation
of $V(x)$ which allows for an exact solution of the equation of motion.
For small $x$ one can derive the asymptotic expansion
\begin{equation}\label{asyexpleft}
 V_1(x) =-\frac{\lambda }{x}-2\lambda \zeta(3)  x^2 +O\left(x^4\right)
 \;,
\end{equation}
using the zeta function $\zeta(z)$, see \cite{NIST}, chapter 25.
The leading term $-\frac{\lambda }{x}$ represents the potential energy due to the left primary mirror
charge. Analogously, in the neighbourhood of the right plate we have
\begin{equation}\label{asyexpright}
 V_1(x) =-\frac{\lambda }{1-x}-2\lambda \zeta (3) (x-1)^2 +O\left((x-1)^4\right)
 \;,
\end{equation}
where $\frac{\lambda }{x-1}$ represents the potential energy due to the right primary mirror charge.
The potential energy, which is only due to the secondary mirror charges will be defined by
\begin{equation}\label{defV2}
 V_2(x) = V_1(x) +\lambda\left(\frac{1}{x}+\frac{1}{1-x}\right)
 \;.
\end{equation}
It is almost constant, see Figure \ref{FIGAPP}.

In the middle between the plates the potential $V_1$ can be better approximated by the Taylor expansion
\begin{equation}\label{taylor}
 V_1(x)=-2 \lambda  \log (4)-14 \lambda
 \zeta(3) \left(x-1/2\right)^2 +O\left((x-1/2)^4\right)
 \;,
\end{equation}
which represents the influence of all mirror charges in the vicinity of the
point $x=1/2$ at which the forces of all mirror charges cancel each other out.
Hence we will choose the following analytical approximation $V_{app}$ of the
partial potential $V_1(x)$
\begin{equation}\label{analapp}
 V_{app}(x) = \left\{
\begin{array}{r@{\quad : \quad}l}
 -\frac{\lambda }{x} &0 < x\le 1/3\;, \\
-2 \lambda  \log (4)-14 \lambda \zeta(3) \left(x-1/2\right)^2 & 1/3 < x <2/3\;,\\
-\frac{\lambda }{1-x} & 2/3\le x < 1 \;.
 \end{array}
 \right.
\end{equation}
For an example of the quality of this approximation see Figure \ref{FIGAPP}.

%%%%%%%%%%%%%%%%%%%%%%%%%%%%%%%%%%%%%%%%%%%%%%%%%%%%%%%%%%%%%%%%%%%%%%%%%%%%%%%%%%%
\subsection{Motion under the influence of a single mirror charge}\label{sec:MC}
%%%%%%%%%%%%%%%%%%%%%%%%%%%%%%%%%%%%%%%%%%%%%%%%%%%%%%%%%%%%%%%%%%%%%%%%%%%%%%%%%%%

The calculations considered in this subsection can be viewed on the one hand
as an approximation of the motion of a point charge in the domain $0<x<1/3$
or $2/3 < x <1$ of the plate capacitor according to
the analytical approximation of $V(x)$ in this domain, see (\ref{analapp}).
On the other hand, these calculations can also be interpreted independently
as the solution of the exact equation of motion of a point charge in the
presence of only one conductive plate.

We will consider the two cases of motion near the left or right plate separately.

%%%%%%%%%%%%%%%%%%%%%%%%%%%%%%%%%%%%%%%%%%%%%%%%%%%%%%%%%%%%%%%%%%%%%%%%%%%%%%%%%%%
\subsubsection{Motion under the influence of the left plate, large energy case}\label{sec:MCL}
%%%%%%%%%%%%%%%%%%%%%%%%%%%%%%%%%%%%%%%%%%%%%%%%%%%%%%%%%%%%%%%%%%%%%%%%%%%%%%%%%%%
Again using dimensionless quantities we can write the potential energy as
\begin{equation}\label{potMCL}
V(x) = -x -\frac{\lambda}{x}
\;.
\end{equation}
As with all one-dimensional mechanical problems with constant total energy $E$,
the inverse function $t(x)$ can be expressed by the indefinite integral
\begin{equation}\label{txint}
 t(x) = \int \frac{dx}{\sqrt{2(E-V(x))}}= \frac{1}{\sqrt{2}}
 \int \frac{dx}{\sqrt{E+x +\frac{\lambda}{x}}}
 = \frac{1}{\sqrt{2}} \int \frac{2 y}{\sqrt{E+y^2 +\frac{\lambda}{y^2}}}\,dy=
 \sqrt{2} \int \frac{y^2}{\sqrt{E y^2+ y^4 +\lambda}}\,dy
 \;,
\end{equation}
where we have substituted $x=y^2$. It is clear that this kind of
expression can be written in terms of elliptical integrals.
Here we will only discuss the large energy case, that is,
\begin{equation}\label{largeE}
 E> 2\sqrt{\lambda}
 \;.
\end{equation}
For the other cases see Appendix \ref{sec:AE}.
In order to transform the integral (\ref{txint}) into some kind of
standard form we factor the polynomial in the square root according to
\begin{eqnarray}\label{factorpolya}
 E y^2+ y^4 +\lambda &= &\left( A^2+y^2\right) \left( B^2+y^22\right),\quad \mbox{where}\\
 \label{factorpolyb}
 A^2&=& \frac{E}{2}-\sqrt{\frac{E^2}{4}-\lambda}\quad \mbox{and}\quad B^2= \frac{E}{2}+\sqrt{\frac{E^2}{4}-\lambda}
 \;.
\end{eqnarray}
Note that $\frac{E^2}{4}-\lambda>0$ by virtue of (\ref{largeE}) and hence $0<A<B$.
Upon the substitution $y=A u$ the integral (\ref{txint}) can be further transformed into
\begin{eqnarray}\label{integrala}
 t&\stackrel{(\ref{factorpolya})}{=} &\sqrt{2} \int \frac{y^2}{\sqrt{\left(A^2+ y^2 \right) \left(B^2+ y^2\right)}}\,dy
 = A^3 \sqrt{2}\int \frac{u^2}{\sqrt{\left(A^2+A^2 u^2 \right) \left(B^2+ A^2 u^2 \right)}}\,du\\
 \label{integralb}
 &=&\frac{A^2}{B} \sqrt{2}\int \frac{u^2}{\sqrt{\left(1+k^2 u^2\right) \left(1+ u^2 \right)}}\,du \quad \mbox{with}\quad
 k^2:=\frac{A^2}{B^2}<1
 \;.
\end{eqnarray}
This integral can be evaluated with the aid of computer algebra software which gives
\begin{eqnarray}\label{integralc}
 t&=&
 -{\sf i}B \sqrt{2} \left(E\left({\sf i} \sinh ^{-1}(u)\left|k^2\right.\right)
 -F\left({\sf i} \sinh ^{-1}(u)\left|k^2\right.\right)\right)\\
 \label{integrald}
&=&-{\sf i}B \sqrt{2}  \left(E\left({\sf i} \sinh ^{-1}\left(\left.\frac{\sqrt{x}}{A}\right)\right|k^2\right)
 -F\left({\sf i} \sinh ^{-1}\left(\left.\frac{\sqrt{x}}{A}\right)\right|k^2\right)\right)
 \;,
\end{eqnarray}
where $F(z|k^2)$ and $E(z|k^2)$ denote the Legendre's elliptic integrals
of the first and second kind, see \cite{NIST}, \S 19, and the above substitutions have been undone.
It is possible to transform (\ref{integralc}) into a manifest real form,
which looks a little more complicated, but this is not necessary because
(\ref{integrald}) can be evaluated numerically without any problems.

%%%%%%%%%%%%%%%%%%%%%%%%%%%%%%%%%%%%%%%%%%%%%%%%%%%%%%%%%%%%%%%%%%%%%%%%%%%%%%%%%%%
\subsubsection{Motion under the influence of the right plate, large energy case}\label{sec:MCR}
%%%%%%%%%%%%%%%%%%%%%%%%%%%%%%%%%%%%%%%%%%%%%%%%%%%%%%%%%%%%%%%%%%%%%%%%%%%%%%%%%%%

The mathematical treatment is largely analogous to that of the preceding subsection.
The potential energy now assumes the form
\begin{equation}\label{potMCR}
V(x) = -x -\frac{\lambda}{1-x}
\;,
\end{equation}
and $t(x)$ is given by the integral
\begin{equation}\label{txintR}
 t(x) = \frac{1}{\sqrt{2}}
 \int \frac{dx}{\sqrt{E+x +\frac{\lambda}{1-x}}}
 = -\frac{1}{\sqrt{2}} \int \frac{2 y}{\sqrt{E+1-y^2 +\frac{\lambda}{y^2}}}\,dy=
 -\sqrt{2} \int \frac{y^2}{\sqrt{(E+1) y^2- y^4 +\lambda}}\,dy
 \;,
\end{equation}
where we have substituted $1-x=y^2$.
We factor the polynomial in the square root according to
\begin{eqnarray}\label{factorpolyRa}
(E+1) y^2- y^4 +\lambda &= &\left( \tilde{A}^2-y^2\right) \left(\tilde{B}^2+y^2\right),\quad \mbox{where}\\
 \label{factorpolb}
 \tilde{A}^2&=& \frac{E+1}{2}+\sqrt{\frac{(E+1)^2}{4}+\lambda}\quad \mbox{and}\quad \tilde{B}^2= -\frac{E+1}{2}+\sqrt{\frac{(E+1)^2}{4}+\lambda}
 \;,
\end{eqnarray}
such that  $0<\tilde{B}<\tilde{A}$ if the large energy condition holds in the slightly weakened form
\begin{equation}\label{lecR}
 E>0
 \;.
\end{equation}

Upon the substitution $y=\tilde{A} u$ the integral (\ref{txintR}) can be further transformed into
\begin{eqnarray}\label{integralRa}
 t&\stackrel{(\ref{factorpolyRa})}{=} & -\sqrt{2} \int \frac{y^2}{\sqrt{\left(\tilde{A}^2- y^2 \right) \left(\tilde{B}^2+ y^2\right)}}\,dy
 = -\tilde{A}^3 \sqrt{2}\int \frac{u^2}{\sqrt{\left(\tilde{A}^2-\tilde{A}^2 u^2 \right) \left(\tilde{B}^2+ \tilde{A}^2 u^2 \right)}}\,du\\
 \label{integralRb}
 &=&-\frac{\tilde{A}^2}{\tilde{B}} \sqrt{2}\int \frac{u^2}{\sqrt{\left(1+\tilde{k}^2 u^2\right) \left(1- u^2 \right)}}\,du \quad \mbox{with}\quad
 \tilde{k}^2:=\frac{\tilde{A}^2}{\tilde{B}^2}>1
 \;.
\end{eqnarray}
This integral is evaluated as
\begin{eqnarray}\label{integralRc}
 t&=& -\tilde{B} \sqrt{2}\left( E\left(\arcsin(u)\left|\right.-k^2\right)-F\left(\arcsin(u)\left|\right.-k^2\right)\right)\\
 &=&-\tilde{B} \sqrt{2}\left( E\left(\arcsin\left(\left.\frac{\sqrt{1-x}}{\tilde{B}}\right)\right|-k^2\right)
 -F\left(\arcsin\left(\left.\frac{\sqrt{1-x}}{\tilde{B}}\right)\right|-k^2\right)\right)
 \;,
\end{eqnarray}
where the above substitutions have been undone.

%%%%%%%%%%%%%%%%%%%%%%%%%%%%%%%%%%%%%%%%%%%%%%%%%%%%%%%%%%%%%%%%%%%%%%%%%%%%%%%%%%%
\subsection{Motion under the influence of both plates}\label{sec:MCM}
%%%%%%%%%%%%%%%%%%%%%%%%%%%%%%%%%%%%%%%%%%%%%%%%%%%%%%%%%%%%%%%%%%%%%%%%%%%%%%%%%%%

\begin{figure}[ht]
  \centering
    \includegraphics[width=0.7\linewidth]{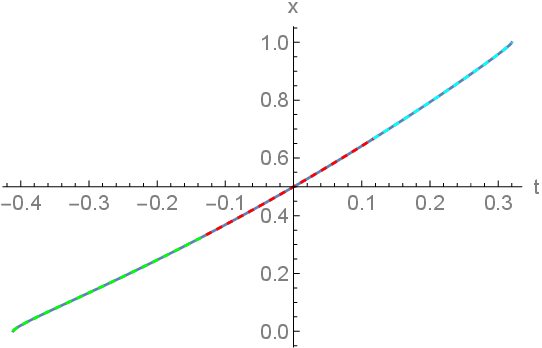}
  \caption{
 We show the numerical integration $x(t)$ of the equation of motion for the point charge
 with the exact potential energy (\ref{totalpot}) (dashed black curve), with the choice
 of parameters $\lambda =0.01$ and $E=0.4$. This result is compared with the analytical approximations
 of the motion corresponding to the intervals $0\le x<1/3$ (green curve), $1/3\le x \le 2/3$ (red curve),
 and $2/3<x\le 1$ (cyan curve), as explained in the main text. The difference between the two types of curves
 is not visible.
  }
  \label{FIGCOMP}
\end{figure}

In the domain $1/3 < x <2/3$ the motion of the point charge will be approximated by a second order Taylor expansion
of the exact potential according to (\ref{analapp}).
The equation of motion can be solved exactly for this approximation,
and the inverse function $t(x)$, which is considered in analogy to the other cases, reads
\begin{equation}\label{txapp2}
 t(x)=\frac{1}{2 \sqrt{\lambda } \sqrt{7 \zeta (3)}}\log \left(2 \sqrt{\lambda } \sqrt{7 \zeta (3)} \sqrt{2 e+\lambda  \log (256)+7 \lambda  (1-2
   x)^2 \zeta (3)+2 x}+14 \lambda  (2 x-1) \zeta (3)+1\right)
 \;.
\end{equation}

%%%%%%%%%%%%%%%%%%%%%%%%%%%%%%%%%%%%%%%%%%%%%%%%%%%%%%%%%%%%%%%%%%%%%%%%%%%%%%%%%%%
\subsection{Comparison of numerical integration and analytical approximation}\label{sec:MCNA}
%%%%%%%%%%%%%%%%%%%%%%%%%%%%%%%%%%%%%%%%%%%%%%%%%%%%%%%%%%%%%%%%%%%%%%%%%%%%%%%%%%%

Finally, we compare the numerical integration of the equation of motion for the point charge
with the exact potential energy (\ref{totalpot}) with the three analytical approximations
obtained so far and find excellent agreement for the choice of parameters  $\lambda =0.01$ and $E=0.4$,
see Figure \ref{FIGCOMP}.
The time $T$ required for the complete motion from $x=0$ to $x=1$ is calculated by numerical
integration as $T_{\scriptstyle{num}}=0.731182$ and by analytical approximations as
$T_{\scriptstyle{aa}}=0.731476$,
which corresponds to a relative deviation of $\approx 4\times 10^{-4}$.
For the $\lambda$-dependence of $T$ see Section \ref{sec:IC}.

%%%%%%%%%%%%%%%%%%%%%%%%%%%%%%%%%%%%%%%%%%%%%%%%%%%%%%%%%%%%%%%%%%%%%%%%%%%%%%%%%%%
\section{On the $\lambda$-dependence of the influence correction}\label{sec:IC}
%%%%%%%%%%%%%%%%%%%%%%%%%%%%%%%%%%%%%%%%%%%%%%%%%%%%%%%%%%%%%%%%%%%%%%%%%%%%%%%%%%%

\begin{figure}[ht]
  \centering
    \includegraphics[width=0.7\linewidth]{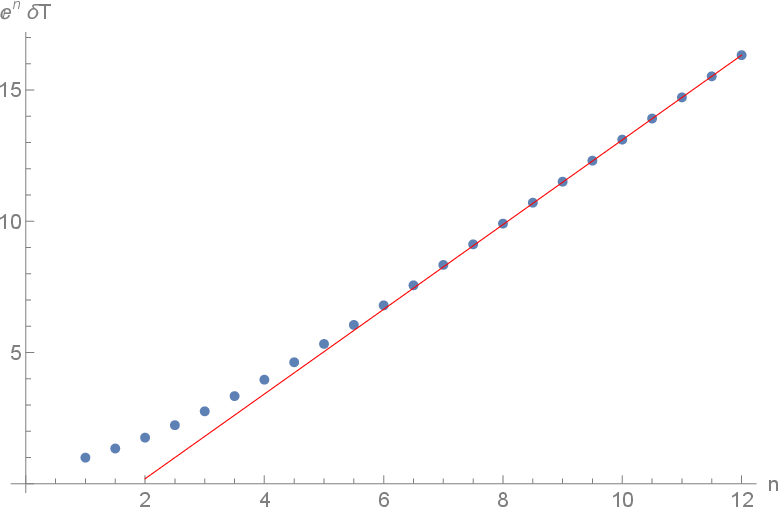}
  \caption{
 Plot of the influence correction $\delta T$ according to (\ref{defic}) divided by $\lambda= {\sf e}^{-n}$
 as a function of $n$. The total energy $E$ has been chosen as $E=0.4$.
 We observe an asymptotically linear dependence for $n\to \infty$, indicated by a red line,
 which suggests that $\delta T = O(-\lambda \log \lambda)$.
  }
  \label{FIGIC}
\end{figure}

Without influence effects, that is, for $\lambda=0$, the time required for the complete motion from $x=0$ to $x=d$
can be elementarily calculated as
\begin{equation}\label{Tele}
 \int_0^d \frac{dx}{\sqrt{\frac{2}{m} \left(E+\frac{U q }{d} x \right)}}
   =
   \frac{\sqrt{2m}\, d \, \left(\sqrt{E+q U}-\sqrt{E}\right)}{q U},
\end{equation}
or, using dimensionless quantities, as
\begin{equation}\label{Teled}
  T(0)=\sqrt{2} \left( \sqrt{E+1}-\sqrt{E} \right)
  \;.
\end{equation}
Taking into account influence effects, this time can be written as
\begin{equation}\label{Telambda}
 T(\lambda) = \int_0^1 \frac{1}{\sqrt{2\left(E-V(x)\right)}} \, dx
 \;,
\end{equation}
with the exact potential energy $V(x)$ given by (\ref{totalpot}).
We define the ``influence correction" by
\begin{equation}\label{defic}
 \delta T :=T(0)- T(\lambda)
 \;.
\end{equation}
and will investigate its $\lambda$-dependence.

Naively, one would think that $\delta T$ goes linearly with $\lambda$ to $0$,
so that $\delta T = O(\lambda)$. However, the numerical results cast doubt on this expectation,
see Figure \ref{FIGIC}, and rather suggest that $\delta T = O(-\lambda \log \lambda)$.

It remains to verify this conjecture analytically.
The obvious attempt to develop the integrand of (\ref{Telambda}) into a
Taylor series with respect to $\lambda$ fails because already the linear term
cannot be integrated over $x$ due to the divergences of the form
$\sim 1/x$ and $\sim 1/(1-x)$.

On the other hand, one can derive from this failure
the hope that the $O(-\lambda \log \lambda)$ behavior of the influence correction
can already be derived for the analytical approximations of $V(x)$,
which was considered in section \ref{sec:MA} for the domains close to $x=0$ and $x=1$,
see (\ref{analapp}). For this purpose, we introduce the influence correction
due to the analytical approximations of $V(x)$
as $\Delta T = \Delta_L T+ \Delta_R T$
and write these terms as integrals using the substitution $x=y^2$ introduced above:
\begin{equation}\label{DeltaT}
\Delta T = \Delta_L T+ \Delta_R T = \int_{0}^{1/2}\left(\frac{y^2}{\sqrt{E+y^2}}- \frac{y^2}{\sqrt{E+y^2+\lambda/y^2}}\right)\,dy +
 \int_{1/2}^{1}\left(\frac{y^2}{\sqrt{E+y^2}}- \frac{y^2}{\sqrt{E+y^2+\lambda/(1-y^2)}}\right)\,dy
 \;.
\end{equation}

\begin{figure}[ht]
  \centering
    \includegraphics[width=1.0\linewidth]{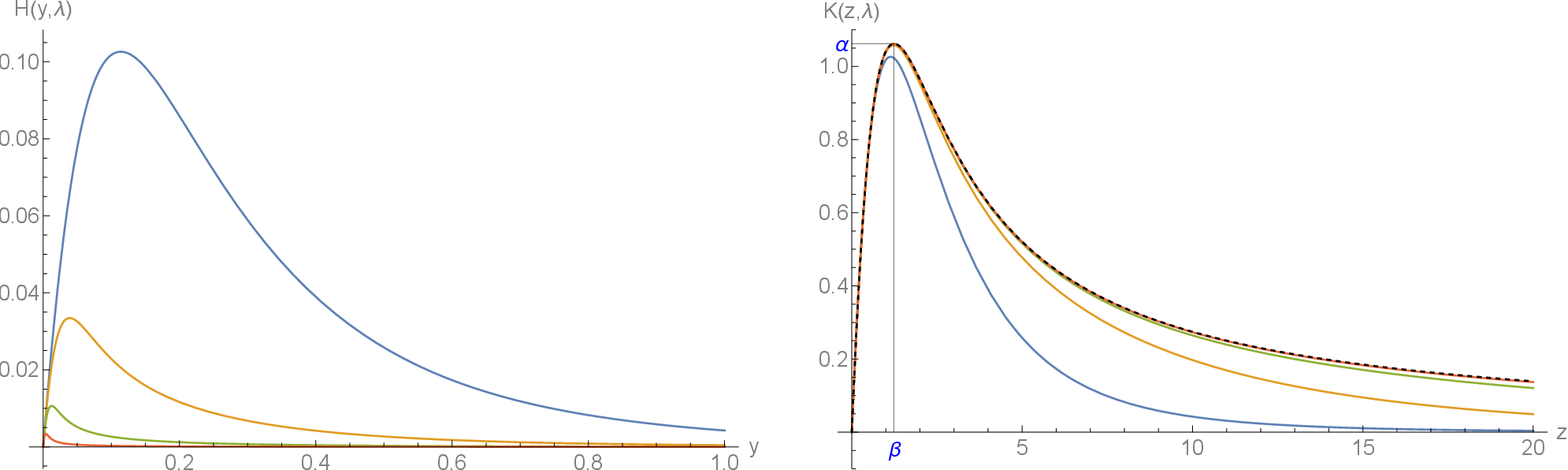}
  \caption{
    Left panel: The family of functions $H(y,\lambda)$ according to (\ref{DetaTL}) for $\lambda=10^{-n}, \, n=2,\ldots,5$,
    starting with the blue curve for $n=2$. The total energy has been chosen as $E=0.4$.\\
    Right panel: The corresponding family of scaled functions $K(z,\lambda)$ according to (\ref{defK}),
    together with the limit function $K(z,0)$ for $n\to \infty$ (dashed black curve). $K(z,0)$
    has its maximum $\alpha=1.06166$ at $z=\beta=1.24301$, in accordance with (\ref{K0max}).
  }
  \label{FIGKS}
\end{figure}

First, let us consider
\begin{equation}\label{DetaTL}
\Delta_L T =  \int_{0}^{1/2}\left(\frac{y^2}{\sqrt{E+y^2}}- \frac{y^2}{\sqrt{E+y^2+\lambda/y^2}}\right)\,dy
=: \int_{0}^{1/2} H(y,\lambda)\, dy
\;.
\end{equation}
The family of functions $H(y,\lambda)$ seems to asymptotically scale with $\sqrt{\lambda}$ w.~r.~t.~argument and function value,
see Figure \ref{FIGKS}. Hence we introduce the scaled family of functions
\begin{equation}\label{defK}
 K(z,\lambda)= \frac{1}{\sqrt{\lambda}} H(\sqrt{\lambda}\,z,\lambda)
 \;,
\end{equation}
such that
\begin{equation}\label{Kint}
 \Delta_L T= \int_{0}^{1/2} H(y,\lambda)\,dy = \lambda \int_{0}^{1/(2\sqrt{\lambda})} K(z,\lambda) \, dz
 \;.
\end{equation}
For $\lambda \to 0$ the function $ K(z,\lambda)$ assumes the limit value
\begin{equation}\label{K0}
 K(z,0)=\sqrt{2} z \left(\frac{1}{\sqrt{E}}-\frac{z}{\sqrt{E z^2+1}}\right)
 \;,
\end{equation}
with a maximum of
\begin{equation}\label{K0max}
 \alpha=K(\beta,0) =\frac{3-\sqrt{5}}{\sqrt{1+\sqrt{5}}\, E} \quad \mbox{at} \quad z=\beta= \sqrt{\frac{\sqrt{5}-E}{2E}}
 \;,
\end{equation}
see Figure \ref{FIGKS}, right panel.
It is possible to explicitly calculate the integral for $\Delta_L T$ using the limit value of $K(z,0)$:
\begin{equation}\label{intDeltaK0}
  \int_{0}^{1/(2\sqrt{\lambda})} K(z,0) \, dz =
  \frac{E-\sqrt{E (E+4 \lambda )}}{4 \sqrt{2} E^{3/2} \lambda }
  +\frac{\log\left(\frac{\sqrt{E}+\sqrt{E+4 \lambda }}{2 \sqrt{\lambda }}\right)}{\sqrt{2} E^{3/2}}
   \;.
\end{equation}
The first term of (\ref{intDeltaK0}) has the series expansion
\begin{equation}\label{intser1}
 \frac{E-\sqrt{E (E+4 \lambda )}}{4 \sqrt{2} E^{3/2} \lambda }=
  -\frac{1}{2 \sqrt{2} E^{3/2}}+\frac{\lambda }{2 \sqrt{2} E^{5/2}}+O\left(\lambda ^2\right)
   \;,
  \end{equation}
and hence does not contribute to the order $O(-\lambda \log \lambda)$ we are looking for.
However, the second term has the asymptotic expansion
\begin{equation}\label{intser2}
\frac{\log\left(\frac{\sqrt{E}+\sqrt{E+4 \lambda }}{2 \sqrt{\lambda }}\right)}{\sqrt{2} E^{3/2}}=
 \frac{\log (E)-\log (\lambda )}{2 \sqrt{2} E^{3/2}}+\frac{\lambda }{\sqrt{2} E^{5/2}}+O\left(\lambda^2\right)
  \;,
\end{equation}
and thus yields, together with (\ref{Kint}), the  desired result  $ \Delta_L T=O(-\lambda \log \lambda)$.

The calculations to prove $ \Delta_R T=O(-\lambda \log \lambda)$ are largely analogous and can be omitted here.

%%%%%%%%%%%%%%%%%%%%%%%%%%%%%%%%%%%%%%%%%%%%%%%%%%%%%%%%%%%%%%%%%%%%%%%%%%%%%%%%%%%%%%%%%%%%%%%%%%%%%%%%%%%%%%%%%%%%%%%%%%%%%%%%%%%%%%%%%%
\section{Summary and Outlook}\label{sec:SO}
%%%%%%%%%%%%%%%%%%%%%%%%%%%%%%%%%%%%%%%%%%%%%%%%%%%%%%%%%%%%%%%%%%%%%%%%%%%%%%%%%%%%%%%%%%%%%%%%%%%%%%%%%%%%%%%%%%%%%%%%%%%%%%%%%%%%%%%%%%

In theoretical physics, it is common to describe the motion of so-called ``test particles"
in external fields in such a way that the influence of the particle
on the sources of the fields is neglected.
An example of this is the motion of a point charge in the field of a plate capacitor,
which is assumed to be constant.
In this case, the neglected influence consists of the formation of induced charges on the capacitor surfaces,
which change the motion of the point charge.

In this example, however, it is also possible to set up an exact equation of motion
taking into account the influence charges.
Although its integration can still only be done numerically, analytical approximations are possible,
which allow an approximate description of the motion using elliptical integrals.
These calculations would probably be practically impossible without the use of computer algebra software.

There are various mathematical and physical aspects that are interesting
beyond the specific problem dealt with. On the one hand, there is the
aforementioned fact that there is a real interaction between the source
and the particle, which can nevertheless be treated exactly.

The Green's function of the plate capacitor, known from the literature
by a double series or mixed series/integral representation using Bessel and trigonometric functions,
could also be obtained in this work by a series representation of the
Coulomb potentials of infinitely many mirror charges. Both representations 
have been shown to agree at the central line passing to the point charge and the mirror charges.
The sum of infinitely many Coulomb potentials of equidistant identical charges diverges; therefore,
for a converging series representation, the Coulomb potentials of pairs
of mirror charges of different signs must be combined. We have thus found
a rare case of only conditional convergence in a physical problem.

For the motion of the point charge, the potential of the mirror charges
at the location of the point charge is required. Remarkably,
the series representation of the potential can be summed up for this special case,
resulting in another application of the psi function in physics.
Other such applications are known from the theory of the Casimir effect, see \cite{E95}, eq.~(5.28),
and from the evaluation of Feynman diagrams, see \cite{C05}.
To a first approximation, the attractive Coulomb potential of the directly neighboring
mirror charge acts in the vicinity of the plates. Although the divergence of this potential
stems from the un-physical limiting case of the fusion of charge and mirror charge,
the strong attraction by the influence charges can certainly be seen as a
rudimentary consideration of the work  of emission that prevents a charge from leaving a
capacitor plate without further ado. Of course, this is only a rough model of the
exact interaction between the point charge and the many-body system
of the capacitor plate, limited by the continuum description of the
macroscopic charge distribution in electrostatics.

In the approximation of the Coulomb potential of the directly neighboring mirror charge,
the motion of the point charge is described by a so-called Cornell potential,
which also appears in the classical Stark effect. The one-dimensional
motion in this potential leads to elliptic integrals. This is an important
special case of motion in the three-dimensional Cornell potential,
which is an integrable problem of classical mechanics
(with the conserved quantities energy, angular momentum around the symmetry axis
and another constant of motion related to the generalized Runge-Lenz vector, see \cite{Cetal95}, \cite{L22}).
We have calculated the motion of the point charge in the Cornell-potential approximation
using elliptic integrals of Legendre's form as well as of Carlson's form.
The latter are a modern variant of elliptic integrals that allow uniform
representations without having to deal with case distinctions.

The ratio of the Coulomb energy of the neighboring mirror charge to the typical
capacitor energy is defined as a dimensionless parameter
$\lambda=\frac{q}{8\pi \epsilon_0 U d}$.
This makes the transit time $T$ from, say, the left to the right plate,
a function of $\lambda$. Contrary to the naive expectation,
$\lambda\mapsto T(\lambda)$ is not analytical at $\lambda=0$, but
$T(\lambda) =T(0) + O(-\lambda \log \lambda)$. Here too,
the singular character of the Coulomb potentials of the primary mirror charges means that
the influence correction behaves differently than expected.

If one inserts the elementary charge for $q$
into the above formula for $\lambda$ and chooses the problem-oriented parameters
$U,d$ of order $1$ in $SI$ units, one gets
$\lambda\sim 10^{-9}$, a very small value. However, if we choose
a macroscopic charge value of, say, $q= 4 \times 10^{-9}\,As$, together with
$d=0.04\,m$, ${\sf m}=0.003\, kg$ and $U=10^3\,V$, we obtain $\lambda \sim 0.45$.
This means that in the experiments mentioned in the Introduction
with an electrostatic pendulum, influence effects cannot be neglected.
For example, if the motion of the point charge has the total energy of
$E = 2\times 10^{-5}\,J$, which is $E=5$ in dimensionless quantities,
we would obtain a transit time  of $T(0)\approx 0.66\, s$ compared with
$T(\lambda)\approx 0.58\,s$, which corresponds to a relative
influence correction of $\delta T/T(0)\approx 12\, \%$.

However, our theory is only suitable for describing the effects on the plates
of the capacitor, not on a conductive sphere that moves between the plates.
Such influence effects could be described in the lowest order by dipole moments
induced by the external electric field. The electric dipole potential leads
to additional influence effects on the plates, which  could be
calculated with an analog method considering an infinite number of mirror dipoles.
This would be a possible extension and application of our work in the future.

\appendix

%%%%%%%%%%%%%%%%%%%%%%%%%%%%%%%%%%%%%%%%%%%%%%%%%%%%%%%%%%%%%%%%%%%%%%%%%%%%%%%%%%%%%%%%%%%%%%%%%%%%%%%%%%%%%%%%%%%%%%%%%%%%%%%%%%%%%%%%%%
\section{Convergence of the series representation of the Green's function}\label{sec:CS}
%%%%%%%%%%%%%%%%%%%%%%%%%%%%%%%%%%%%%%%%%%%%%%%%%%%%%%%%%%%%%%%%%%%%%%%%%%%%%%%%%%%%%%%%%%%%%%%%%%%%%%%%%%%%%%%%%%%%%%%%%%%%%%%%%%%%%%%%%%

In order to investigate the convergence of the series representation of the putative (Dirichlet) Green's function
(\ref{Phi}) it is sensible to split off the three terms $\Phi_0({\mathbf r},{\mathbf r}_0)$ corresponding to the
Coulomb potential of the point charge and of the two primary mirror charges.
This part can be considered as a distribution satisfying the Poisson equation
\begin{equation}\label{PoissonPhi0}
\Delta \Phi_0\left({\mathbf r},{\mathbf r}_0 \right)=
\Delta \Phi\left({\mathbf r},{\mathbf r}_0 \right)=
-\frac{q}{\epsilon_0}\,\delta({\mathbf r}-{\mathbf r}_0)
\end{equation}
and can be regularized by cancelling two Coulomb potentials in the case of $x=0$
(when the point charge hits the left plate and coincides with the left primary mirror charge)
or $x=d$ (analogously for the right plate). Then it will be appropriate to postulate (\ref{PoissonPhi0})
only for the open domain $0<x<d$ between the plates.

After having subtracted  $\Phi_0\left({\mathbf r},{\mathbf r}_0\right)$ from the series representation
(\ref{Phi}) we consider only the remaining contributions from the first generation of mirror charges, i.~e.~,
\begin{equation}\label{Phi1}
\Phi_1\left({\mathbf r},{\mathbf r}_0\right)= \frac{q}{4\pi \epsilon_0}\sum_{n=2}^\infty
 \frac{(-1)^n}{\left|{\mathbf r}-{\mathbf R}_1(n) \right|}
 \;,
\end{equation}
such that $-d/2 \le \xi, \xi_0 \le d/2$ for ${\mathbf r}=(\xi,\eta,\zeta)$ and ${\mathbf r}_0=(\xi_0,0,0)=(x-d/2,0,0)$.
The contribution from the second generation can be treated completely analogously.

\begin{figure}[ht]
  \centering
    \includegraphics[width=0.7\linewidth]{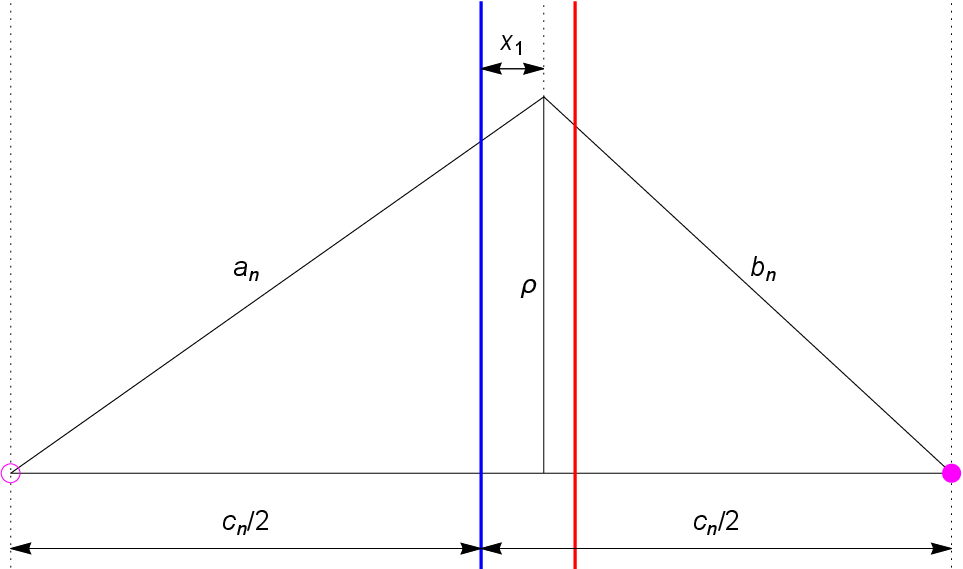}
  \caption{
  Sketch of the geometry underlying the contribution (\ref{pair}) of a pair of mirror charges to the Green's function
  $ \Phi({\mathbf r},{\mathbf r}_0)$. The plates of the capacitor are indicated by blue and red vertical lines.
  We display the triangle with the edges  $a_n, b_n$ and $c_n$
  spanned by two mirror charges with numbers $n, n+1$, $n$ even, of the first generation (magenta dot and circle) and the
  point with coordinates ${\mathbf r}=(-d/2+x_1,\eta,\zeta)$, where the potential is evaluated. The height
  of the triangle is the cylinder coordinate  $\rho=\sqrt{\eta^2+\zeta^2}$. Note that $b_n\le a_n$
  always applies because the left capacitor plate (blue line) lies exactly in the middle between the mirror charges.
  }
  \label{FIGDK}
\end{figure}

It is obvious that the series (\ref{Phi1}) does not converge unconditionally, i.~e.,
independent of all rearrangements of the members of the series,
since the partial series for the odd and even $n$ diverge in the same way
as the harmonic series $\sum_{n=1}^\infty \frac{1}{n}$.
However, convergence can be achieved if we group together
the $n$-th and the $(n+1)$-th term of (\ref{Phi1}) for even $n$, having different signs.
This corresponds to a mirror charge $q$ with coordinate $X_1(n)=n d +x - d/2> d/2$ and its mirror $-q$ with
coordinate $X_1(n+1)=-n d -x -d/2< -d/2$.
We consider the contribution to (\ref{Phi1}) corresponding to this pair of mirror charges
(ignoring the pre-factor $\frac{q}{4\pi \epsilon_0}$):
\begin{equation}\label{pair}
\phi_n:=  \frac{1}{\left|{\mathbf r}-{\mathbf R}_1(n) \right|}- \frac{1}{\left|{\mathbf r}-{\mathbf R}_1(n+1) \right|}\;,
\quad n \mbox{  even}
\;.
\end{equation}
We use the notation explained in Figure \ref{FIGDK} and write
\begin{equation}\label{anbn}
 \phi_n = \frac{1}{b_n} - \frac{1}{a_n} = \frac{a_n-b_n}{a_n\,b_n} =\frac{(a_n-b_n)(a_n+b_n)}{a_n\,b_n\,(a_n+b_n)} =
 \frac{a_n^2-b_n^2}{a_n\,b_n\,(a_n+b_n)} \ge 0
 \;,
\end{equation}
since $a_n\ge b_n$ always applies and $a_n=b_n$ only for $x=0$, see Figure \ref{FIGDK}.
This term can be further evaluated with the help of Pythagoras' theorem which yields
\begin{equation}\label{an2bn2}
 a_n^2= \rho^2 +\left(\frac{c_n}{2}+x_1\right)^2, \quad \mbox{and}\quad b_n^2= \rho^2 +\left(\frac{c_n}{2}-x_1\right)^2
 \;,
\end{equation}
see Figure \ref{FIGDK}, and hence
\begin{equation}\label{bn2man2}
 a_n^2-b_n^2 = 2\, c_n\, x_1
 \;.
\end{equation}
We conclude
\begin{eqnarray}
\label{ab1}
 \phi_n & =& \frac{1}{b_n}-  \frac{1}{a_n} \stackrel{(\ref{anbn},\ref{bn2man2})}{=}
 \frac{2 c_n x_1}{a_n b_n (a_n+b_n)}
   \le \frac{2 x_1}{a_n b_n}\le \frac{2 d}{a_n b_n}
   \stackrel{(\ref{an2bn2})}{\le} \frac{2 d}{\left(\frac{c_n}{2}+x_1\right) \left(\frac{c_n}{2}-x_1\right)} \\
   \label{ab2}
   &= & \frac{2d}{\frac{1}{4} c_n^2 -x_1^2}
   \le \frac{8 d}{c_n^2-4 d^2}=\frac{8 d}{\left(2 d n +x \right)^2-4 d^2}
   \le \frac{8 d}{4 d^2  n^2-4 d^2}=\frac{2}{d \left(n^2 -1 \right)} < \frac{2}{d (n-1)^2}
   \;.
\end{eqnarray}
Here we have used in line (\ref{ab1}) that, due to the triangle inequality,  $c_n\le a_n+ b_n$
and that $0\le x_1 \le d$ due to the restriction of ${\mathbf r}$ to the space between the plates.
In the line  (\ref{ab2}) we have further used the equation
\begin{equation}\label{cn}
 c_n=X_1(n)- X_1(n+1)=\left(n d +x - d/2\right) - \left(-n d -x -d/2 \right) = 2\,d\,n+x \ge 2\,d\,n
 \;,
\end{equation}
and that $n^2 -1 > (n-1)^2$ for $n\ge 2$.

Summarizing, we have found that the series $\sum_{n=2,4,6,\ldots}^\infty \phi_n$
is majorized by the convergent series
$\frac{2}{d } \sum_{n=2,4,6\ldots}^\infty \frac{1}{(n-1)^2}=\frac{\pi^2}{4d}$.
Hence also $\Phi_1\left({\mathbf r},{\mathbf r}_0\right)$ converges
w.~r.~t.~the arrangement of the series according to the pairs  $\phi_n$.
Since the upper bound in (\ref{ab2}) does not depend on ${\mathbf r}$
or  ${\mathbf r}_0$  the convergence is uniform in the region
$-d/2 \le \xi, \xi_0 \le d/2$.

In order to verify the Poisson equation (\ref{PoissonPhi0}) we have to show
\begin{equation}\label{PoissonPhi1}
 \Delta \Phi_1 = \Delta \left( \Phi-\Phi_0\right) = 0,\quad \mbox{for}\quad  -\frac{d}{2} < \xi <\frac{d}{2}
 \;.
\end{equation}
This follows because that the sources of
the potential $\Phi_1$ consist of an infinite number of
secondary mirror charges outside the space between the plates.

%%%%%%%%%%%%%%%%%%%%%%%%%%%%%%%%%%%%%%%%%%%%%%%%%%%%%%%%%%%%%%%%%%%%%%%%%%%%%%%%%%%%%%%%%%%%%%%%%%%%%%%%%%%%%%%%%%%%%%%%%%%%%%%%%%%%%%%%%%
\section{Comparison with other representations of the Green's function}\label{sec:CO}
%%%%%%%%%%%%%%%%%%%%%%%%%%%%%%%%%%%%%%%%%%%%%%%%%%%%%%%%%%%%%%%%%%%%%%%%%%%%%%%%%%%%%%%%%%%%%%%%%%%%%%%%%%%%%%%%%%%%%%%%%%%%%%%%%%%%%%%%%%
As mentioned in the Introduction there exist other series representations of the Green's function
$G({\mathbf r},{\mathbf r}')$
of the plate capacitor, especially in \cite{J21}, problem 3.17a and 3.17b. The latter is a mixed
series/integral representation. It will be instructive to compare these with our explicit
form (\ref{phi}) obtained for the special case where both, ${\mathbf r}$ and ${\mathbf r}'$, are chosen to lie
on the $\xi$-axis between the plates.

We first consider the double series representation of $G({\mathbf r},{\mathbf r}')$ according to \cite{J21}, problem 3.17a:
\begin{equation}\label{J317a}
 G({\mathbf r},{\mathbf r}') = \frac{4}{L} \sum_{n=1}^{\infty}\sum_{m=-\infty}^{\infty}
 {\sf e}^{{\sf i}m(\phi- \phi')}\sin \left(\frac{n \pi z}{L} \right)\sin \left(\frac{n \pi z'}{L} \right)
 I_m\left(\frac{n \pi}{L}\rho_< \right) K_m\left(\frac{n \pi}{L}\rho_> \right)
 \;.
\end{equation}
Here $z,\rho,\phi$ are cylindrical coordinates of ${\mathbf r}$, analogously for $z',\rho',\phi'$
and the two plates of the capacitor lie at $z=0$ and $z=L$. (Hence the $z$-axis corresponds
to our $\xi$-axis and the $L$ to our $d$.) The $I_m(z)$ and $K_m(z)$ denote the modified
Bessel functions of the 1st resp.~2nd kind. 
$\rho_<$ denotes the minimum of $\rho$ and $\rho'$
and $\rho_>$ the corresponding maximum. If both position vectors 
${\mathbf r}$ and ${\mathbf r}'$ are located at the $z$-axis we have $\rho=\rho'=0$ and hence $\rho_<=\rho_>=0$.
Due to $I_m(0)=0$ for $m\neq 0$ and $I_0(0)=1$ only the $m=0$-term of the sum survives
and the term ${\sf e}^{{\sf i}m(\phi- \phi')}$ can be replaced by $1$. This is highly 
plausible due to the azimuthal symmetry of the problem. However, the term
$K_0\left(\frac{n \pi}{L}\rho_> \right)$ diverges for $\rho_>\to 0$ and hence
the series representation (\ref{J317a}) is not suitable for comparison with (\ref{phi}),
at least if we adopt the obvious choice and identify the central line with the $z$-axis.
It may be, nevertheless, useful for evaluating the Green's function outside the $z$-axis, see below.

Next we consider the mixed series/integral representation of 
$G({\mathbf r},{\mathbf r}')$ according to \cite{J21}, problem 3.17b:
\begin{equation}\label{J317b}
  G({\mathbf r},{\mathbf r}') = 2 \sum_{m=-\infty}^{\infty} \int_{0}^{\infty} dk\,
 {\sf e}^{{\sf i}m(\phi- \phi')} J_m(k \rho) J_m(k \rho') 
 \frac{\sinh(k z_<)\sinh(k (L-z_>))}{\sinh(k L)}
 \;,
\end{equation}
where the $J_m(z)$ denote the Bessel functions and $z_<$, resp.~$z_>$, the 
minimum, resp.~maximum, of $z$ and $z'$. 
Again, we restrict ourselves to the case $\rho=\rho'=0$ and hence the sum shrinks to the term with $m=0$. 
The remaining integral can be expressed in terms of the hypergeometric functions 
$_2F_1(a,b,c;z)$, as far as the indefinite form is considered, and the definite
form can be simplified to 
\begin{equation}\label{J317bsimp}
  G({\mathbf r},{\mathbf r}') =
  \frac{1}{2L} \left(\psi\left(1-\frac{z_<+z_>}{2L}\right)-
  \psi\left(1+\frac{z_<-z_>}{2L}\right)-
  \psi\left(\frac{z_>-z_<}{2L}\right)+
  \psi\left(\frac{z_<+z_>}{2L}\right)\right)
 \;.
\end{equation}
This expression can be shown to agree with (\ref{phi}) for both cases,
$z\le z'$ and $z\ge z'$, 
by using the equations
$z=\xi+L/2,\; z'=\xi_0+L/2,\;L=d$ and $\psi(z+1)=\psi(z)+1/z$. For the latter see 
\cite{NIST}, eq.~5.5.2.

Finally, we want to use the series representation (\ref{J317a}) 
to calculate the total influenced charge on the left plate. 
To do this, we choose $\rho=0$ again,
differentiate the resulting expression w.~r.~t.~$z'$ and set $z'=0$. This gives
essentially the normal component of the electric field at the left hand plate and is hence proportional
to the surface density $\sigma_1$:
\begin{equation}\label{sigma1}
 \sigma_1 = - \frac{q}{4\pi} \frac{\partial}{\partial z'}\left. G({\mathbf r},{\mathbf r}') \right|_{z'=0}
 = - \frac{q}{\pi L}\sum_{n=1}^{\infty} \sin\frac{n \pi z}{L}\,\left(\frac{n \pi}{L}\underbrace{\cos \frac{n \pi z'}{L}}_{1} \right)
 K_0\left( \frac{n \pi \rho}{L}\right)=- \frac{q}{L^2}\sum_{n=1}^{\infty} n \sin\frac{n \pi z}{L}\, 
 K_0\left( \frac{n \pi \rho}{L}\right)
 \;.
\end{equation}
The total influenced charge $Q_1$ on the left hand plate is obtained by the integral
\begin{equation}\label{totalcharge1}
 Q_1=\int_{0}^{\infty} \sigma_1\, 2\pi \rho\, d\rho
 \;.
\end{equation}
In a sub-calculation we obtain
\begin{equation}\label{Kint}
 \int_{0}^{\infty} K_0\left( \frac{n \pi \rho}{L}\right)\, 2\pi \rho\, d\rho =
  \frac{2 \pi L^2}{n^2 \pi^2}\underbrace{\int_{0}^{\infty}K_0(r) r dr}_{1}= \frac{2L^2}{n^2 \pi}
  \;,
\end{equation}
using the substitution $ r=\frac{n \pi \rho}{L}$ and \cite{NIST}, 10.43.19.
Insering this result into (\ref{totalcharge1}) we finally obtain
\begin{equation}\label{totalcharge1a}
 Q_1 = - \frac{q}{L^2}\frac{2 L^2}{\pi}\sum_{n=1}^{\infty} \frac{1}{n} \sin\frac{n \pi z}{L} 
 = -q\, \left(1-\frac{z}{L} \right)
 \;,
\end{equation}
using the Fourier series of the sawtooth wave \cite{W22}.
By an analogous calculation we obtain for the influenced charge $Q_2$ on the right hand plate
\begin{equation}\label{totalcharge1a}
 Q_2=  -q\, \frac{z}{L} 
 \;,
\end{equation}
such that $Q_1+Q_2=-q$. These results agree with \cite{J21}, problem 1.13, 
where  a solution using Green's reciprocity theorem is suggested. 
A calculation using the set of mirror charges for the plate capacitor is much more difficult, 
see \cite{N82}.

%%%%%%%%%%%%%%%%%%%%%%%%%%%%%%%%%%%%%%%%%%%%%%%%%%%%%%%%%%%%%%%%%%%%%%%%%%%%%%%%%%%%%%%%%%%%%%%%%%%%%%%%%%%%%%%%%%%%%%%%%%%%%%%%%%%%%%%%%%
\section{Motion under the influence of the left plate, arbitrary energy}\label{sec:AE}
%%%%%%%%%%%%%%%%%%%%%%%%%%%%%%%%%%%%%%%%%%%%%%%%%%%%%%%%%%%%%%%%%%%%%%%%%%%%%%%%%%%%%%%%%%%%%%%%%%%%%%%%%%%%%%%%%%%%%%%%%%%%%%%%%%%%%%%%%%

\begin{figure}[ht]
  \centering
    \includegraphics[width=0.7\linewidth]{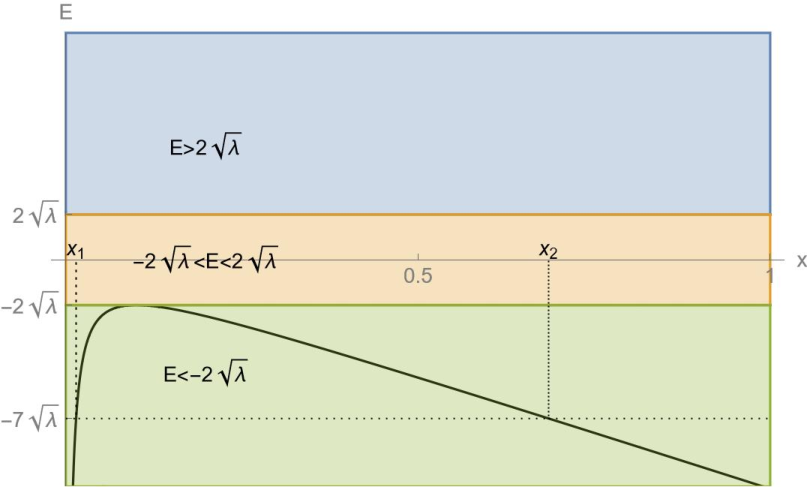}
  \caption{
  Sketch of various cases for the evaluation of the integral (\ref{txint}). The parameter $\lambda$ is chosen as $\lambda=0.01$.
  The potential energy $V(x) = -x -\lambda/x$ (black curve) has its maximum $E_{\scriptstyle{max}}=-2\sqrt{\lambda}$
  at $x=\sqrt{\lambda}$. For $E<-2\sqrt{\lambda}$ there are two turning points
  $x_1=\frac{1}{2} \left(-E-\sqrt{E^2-4 \lambda }\right) < x_2=\frac{1}{2} \left(-E+\sqrt{E^2-4 \lambda }\right)$
  where $E=V(x_i),\, i=1,2,$ such that the point charge cannot enter the range $x_1<x<x_2$.
  We have indicated the example of $E=-7 \sqrt{\lambda}$ by dotted black lines.
  }
  \label{FIGRP}
\end{figure}

\begin{figure}[ht]
  \centering
    \includegraphics[width=1.0\linewidth]{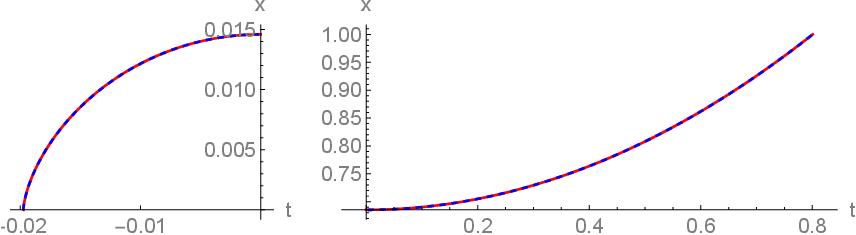}
  \caption{
  Comparison of the motion $x(t)$ of the point charge calculated analytically according to (\ref{19.29.20}) (blue dashed curves)
  with the numerical results (red curves). We have chosen $\lambda=0.01$ and $E=-0.7$.  There are two turning points
  $x_1=0.0145898$ and $x_2=0.68541$ where the velocity of the point charge vanishes, corresponding to the example given in Figure \ref{FIGRP}.
  Left panel: Range $0\le x\le x_1$. Right panel: Range $x_2\le x \le 1$.
  }
  \label{FIGPP}
\end{figure}

\begin{figure}[ht]
  \centering
    \includegraphics[width=1.0\linewidth]{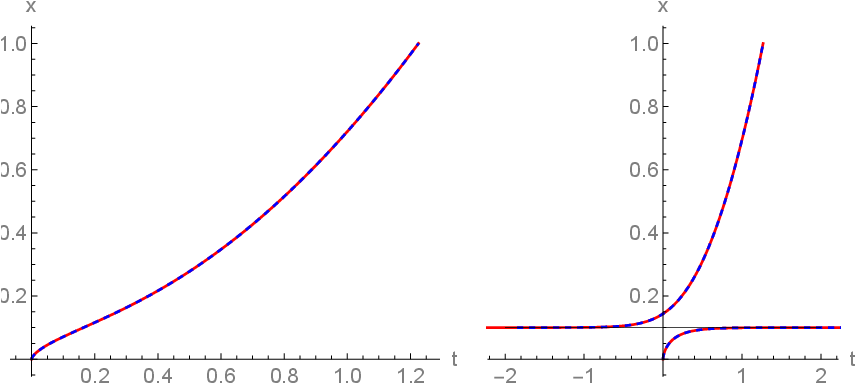}
  \caption{
  Comparison of the motion $x(t)$ of the point charge calculated analytically according to (\ref{19.29.20}), resp.,
 (\ref {motioni0}) (blue dashed curves)
  with the numerical results (red curves). We have chosen $\lambda=0.01$.
  Left panel: Case  $E=-0.5$ with complex $a_j,\;j=1,2$. Right panel: Case  $E=-2\sqrt{\lambda}=-0.2$.
  The two branches of the plot correspond to the motion approaching or leaving the unstable rest position
  of $x=\sqrt{\lambda}=0.1$ where $V(x)$ has a global maximum for $t \to \pm\infty$.
  }
  \label{FIGQQ}
\end{figure}

If we consider the motion of a point charge in a potential $V(x) = -x -\lambda/x$ for arbitrary energies $E$,
it is more convenient to use the Carlson symmetric forms of elliptic integrals, see \cite{NIST}, section 19,
and \cite{C64}, instead of the Legendre form.
We will quote the corresponding equations (19.29.20), (19.29.22), (19.16.5) in \cite{NIST}, adapted to our notation:
\begin{equation}\label{19.29.20}
 \int_{y_0}^{y1} \frac{y^2 dy}{\sqrt{Q_1(y)\,Q_2(y)}} =
 \frac{1}{3} a_1 a_2 R_D\left(U^2+a_1 b_2,U^2+a_2 b_1,U^2 \right) +\frac{y_0 y_1}{U}
 \;,
\end{equation}
where
\begin{equation}\label{Qj}
 Q_j(y)=a_j  + b_j y^2, \, j=1,2
 \;,
\end{equation}
\begin{equation}\label{19.29.22}
 (y_1^2-y_0^2) U=y_1\sqrt{Q_1(y_0)Q_2(y_0)}+y_0 \sqrt{Q_1(y_1)Q_2(y_1)}
 \;,
\end{equation}
and
\begin{equation}\label{19.16.5}
R_D(x,y,z)=\frac{3}{2}\int_{0}^{\infty}\frac{dt}{\sqrt{(t+x)(t+y)(t+z)}\,(t+z)}
\;.
\end{equation}
These equations are especially suited to express the integral  (\ref{txint}) in terms
of the standard Carlson elliptic integral $R_D(x,y,z)$
(a degenerate case of the symmetric Carlson elliptic integral $R_J$ of the second kind)
for all choices of the energy $E$
(except for the special value $E=-2\sqrt{\lambda}$, see below).
In our case we have, see (\ref{factorpolyb}),
\begin{eqnarray}
\label{a1}
  a_1 &=& \frac{E}{2}-\sqrt{\frac{E^2}{4}-\lambda},\quad b_1=1\;, \\
  \label{a2}
  a_2 &=& \frac{E}{2}+\sqrt{\frac{E^2}{4}-\lambda},\quad b_2=1
  \;,
\end{eqnarray}
and
\begin{equation}\label{U}
 U=\frac{y_0 \sqrt{E y_1^2+\lambda +y_1^4}+y_1 \sqrt{E y_0^2+\lambda +y_0^4}}{y_1^2-y_0^2}
 \;.
\end{equation}
In the range $-2\sqrt{\lambda}< E <2\sqrt{\lambda}$, see Figure \ref{FIGRP},
the parameters $a_1$ and $a_2$ will be complex, but the equation (\ref{19.29.20})
can nevertheless be used to calculate the integral (\ref{txint}), see Figure \ref{FIGQQ}, left panel.

We will compare the analytical form of the motion of the point charge with the results from numerical
integration for two different values of $E$, see the Figures \ref{FIGPP} and \ref{FIGQQ}, left panel.
The agreement is perfect.

In the special case of $E=-2\sqrt{\lambda}$ the two polynomials $Q_1(y)$
and $Q_2(y)$ coincide and the integral (\ref{txint}) can be evaluated
in a simpler way without using elliptic integrals.
This case is comparable to the motion of a mathematical pendulum
which reaches its unstable resting position for $t\to \pm\infty$.
For $E=-2\sqrt{\lambda}$ we obtain
\begin{equation}\label{motioni0}
t(x) =\frac{\sqrt{2} \left(x-\sqrt{\lambda }\right) \left(\sqrt{x}-\frac{1}{4} \sqrt[4]{\lambda } \log
   \left(\frac{\left(\sqrt[4]{\lambda }+\sqrt{x}\right)^2}{\left(\sqrt{x}-\sqrt[4]{\lambda
   }\right)^2}\right)\right)}{\sqrt{x} \sqrt{-2 \sqrt{\lambda }+\frac{\lambda }{x}+x}}
   \,.
\end{equation}
For a comparison of this solution with numerical results see Figure \ref{FIGQQ}, right panel.

The case of the motion under the influence of the right plate and arbitrary energy
can be analogously treated using the Carlson symmetric forms of elliptic integrals
and will not be considered  here.

\end{document}